\documentclass[aip,rsi,reprint,a4paper]{revtex4-1}

\usepackage{amssymb,amsmath}
\usepackage{graphicx}
\usepackage{longtable}

\begin{document}

\newcommand{\degree}{$^\circ$}
\newcommand{\note}[1]{\textbf{[#1]}}

\newcommand{\affMBI}{Max Born Institute, Max-Born-Stra{\ss}e 2A, 12489
Berlin, Germany.}
\newcommand{\affLUXOR}{National Research Council - Institute
of Photonics and Nanotechnologies (CNR-IFN), via Trasea 7, I-35131
Padova, Italy.}

\title{Alignment and characterization of the two-stage time delay 
compensating XUV monochromator}

\author{Martin Eckstein}
\author{Johan Hummert}
\affiliation{\affMBI}

\author{Markus Kubin}
\altaffiliation{present address: Helmholtz-Zentrum Berlin, 
Albert-Einstein-Straße 15, 12489 Berlin, Germany.}
\affiliation{\affMBI}

\author{Chung-Hsin Yang}
\altaffiliation{present address: Academia Sinica, 128 Academia Road, Section 
2, Nankang, Taipei 115, Taiwan.}
\affiliation{\affMBI}

\author{Fabio Frassetto}
\author{Luca Poletto}
\affiliation{\affLUXOR}

\author{Marc J. J. Vrakking}
\affiliation{\affMBI}

\author{Oleg Kornilov}
\email{kornilov@mbi-berlin.de}
\affiliation{\affMBI}

\begin{abstract}
We present the design, implementation and alignment procedure for a
two-stage time delay compensating monochromator. 
 %recently employed in
 %time-resolved photoionization studies of nitrogen molecules [M. Eckstein et
 %al, J. Phys. Chem.  Lett. \textbf{6}, 419 (2015)]. 
 The setup spectrally filters the radiation of a high-order harmonic
generation source providing wavelength-selected XUV pulses with a bandwidth
of 300 to 600~meV in the photon energy range of 3 to 50~eV. XUV pulses as
short as $12\pm3$~fs are demonstrated. Transmission of the 400~nm (3.1~eV)
light facilitates precise alignment of the monochromator. This alignment
strategy together with the stable mechanical design of the motorized
beamline components enables us to automatically scan the XUV photon energy
in pump-probe experiments that require XUV beam pointing stability. The
performance of the beamline is demonstrated by the generation of IR-assisted
sidebands in XUV photoionization of argon atoms. 
\end{abstract}

\maketitle

%%%%%%%%%%%%%%%%%%%%%%%%%%%%%%%%%%%%%%%%%%%%%%%%%%%%%%%%%%%%%%%%%%%%%%%%
\section{Introduction}

Ultrafast laser pulses in the extreme-ultraviolet (XUV) range become more
and more instrumental in studies of molecules, liquids and solids. 
Development of novel XUV laser sources such as free electron lasers and
high-order harmonic generation (HHG) sources allows experiments with femtosecond
and even attosecond time resolution, which are capable of following nuclear
and electron dynamics in real time \cite{Calegari2014,Eckstein2015}. 
In many cases ultrafast XUV studies benefit from control of the XUV photon
energy.  Such control is routinely available for free electron lasers.
However, in the HHG process a wide range of XUV energies is inherently
generated and spectral filtering is necessary to produce bandwidth-limited
pulses that are suitable for experiments with simultaneous time and energy
resolution.

Selection of the photon energy from an HHG source requires an XUV 
monochromator. Some of the simplest monochromators are realized using 
specially designed multilayer mirrors optimized for reflection in a 
desired XUV photon energy range \cite{Siffalovic2001}.  This solution is 
convenient when a fixed range of photon energies is needed for the 
experiment, but changing the photon energy requires replacement of the 
multilayer mirrors.

%Moreover
%properties of materials typically used in fabrication of the multilayer
%mirrors limit the photon energy to the range from 25 to ??  eV \cite{ML2}.

More versatile tools are based on diffraction. Such systems have been
realized using spherical and toroidal diffraction gratings
\cite{Nugent-Glandorf2002,Ito2010}, plane gratings in normal and off-plane
configurations \cite{Poletto2007, Dakovski2010, Frietsch2013} and Fresnel
zone plates \cite{Gaudin2008, Ibek2013, Metje2014}.  Designs based on
diffraction gratings can provide very high spectral resolution at the
expense of temporal resolution. Diffraction on a transmission or reflection
grating leads to an optical path difference (OPD) across the spatial profile
of the beam. The path difference is proportional to the number of
illuminated grooves $N$ and can be expressed as $\mathrm{OPD}=Nm\lambda$,
where $m$ is the diffraction order and $\lambda$ is the wavelength of the
light. Correspondingly, there is a delay that accumulates across the
beam profile, which stretches the pulse in time and reduces the temporal
resolution of a pump-probe experiment.

This problem can be overcome by using a time delay compensating scheme with
two diffraction gratings \cite{Villoresi1999,Poletto2004}.  The pulse
stretched by the first diffraction grating is compressed upon reflection
from the second identical grating mounted in a mirrored configuration.  This
optical design has been previously realized by Poletto et al
\cite{Poletto2007} and demonstrated excellent temporal resolution for a
typical HHG source.  Here we describe a new implementation of the time delay
compensating monochromator (TDCM) beamline employing two diffraction
gratings.  The optical design of the monochromator is similar to that
described earlier \cite{Poletto2007} and is also based on off-plane grating
mounts.  The main design goals of the present setup, making it different
from those published earlier, is optimization for a wide range of photon
energies (3 to 50~eV), the accomplishment of stable long-term operation
(days of unattended operation) and reliability in automated switching
between photon energies in the course of experiments.

The next section describes the optical design of the monochromator and lists
parameters of the optical elements used. The third section presents the
mechanical implementation and alignment procedures necessary for stable and
reliable operation.  Section IV is devoted to energy, transmission and pulse
duration characterization of the monochromator setup.  In the last section
application of the setup for the generation of IR-assisted sideband
structures \cite{Kroll1973,Maquet2007} in XUV photoionization of argon
atoms is discussed.

\section{Optical design}

\begin{figure*}[!t]
\centering
\includegraphics[width=\textwidth]{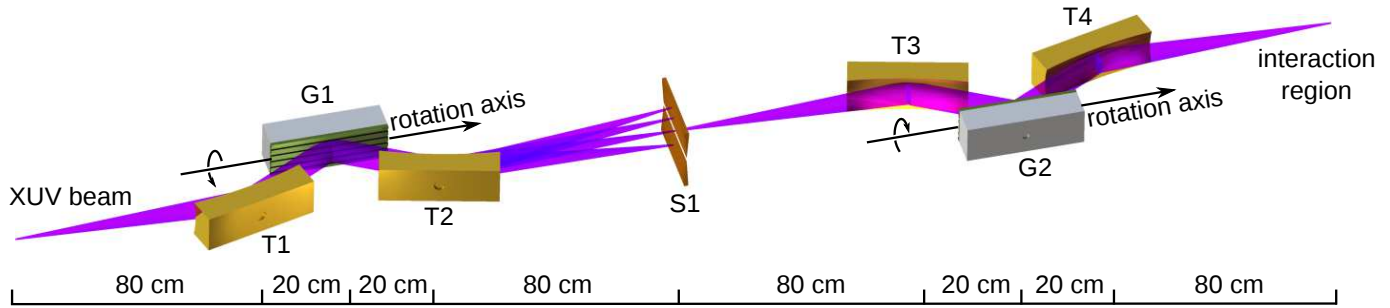}
\caption{\label{fig:mono_layout}
Optical layout of the TDCM, consisting of four toroidal mirrors (T1-T4), two
diffraction gratings (G1, G2) operated in the off-plane mount and the
monochromator slit (S1). The length scale is shown at the bottom. The XUV
beam is spectrally dispersed by the first stage (T1+G1+T2) and is filtered
by the slit (S1). At this this point the beam is stretched in time. The second
stage (T3+G2+T4) recompresses the XUV pulse and guides the beam towards the
interaction region of an experimental endstation.
}
\end{figure*}

A brief description of the monochromator beamline has recently been
published elsewhere \cite{Eckstein2015}. The full optical layout of the
two-grating time delay compensating monochromator (TDCM) is shown in Fig.
\ref{fig:mono_layout}. It consists of four toroidal mirrors, two diffraction
gratings and an exchangeable slit. All mirrors have the same focal length of
80~cm and therefore each pair of toroidal mirrors together with the grating
between them (T1+G1+T2 and T3+G2+T4) forms a Rowland circle configuration,
which minimizes geometric aberrations in beam imaging.

\begin{figure}[ht]
\centering
\includegraphics[width=8.5cm]{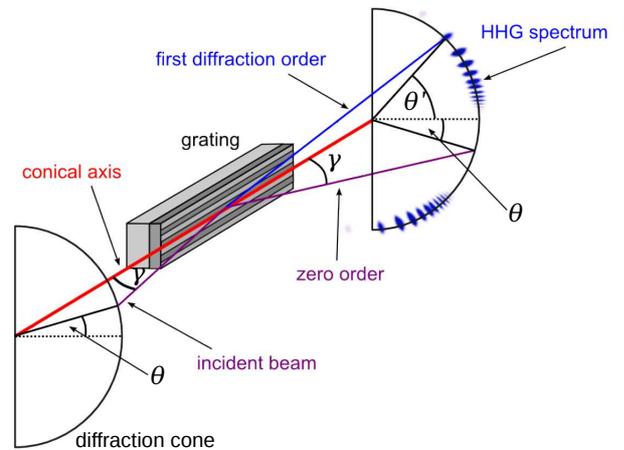}
\caption{\label{fig:con_dif}
Off-plane diffraction mount. The XUV beam is incident on the grating mounted 
with its grooves parallel to the long side of the grating. The diffracted 
beams are located on a cone with the opening half-angle of $\gamma$, which is 
defined as shown in the figure. The angles of the incident 
beam $\theta$ and the diffracted beam $\theta'$ for the given 
wavelength $\lambda$ are connected by Eq. (\ref{eq:con}).
}	
\end{figure}

The first toroidal mirror T1 collimates the XUV beam coming from an HHG
source and reflects it towards the first grating G1. The desired photon
energy is selected by rotating the grating G1 around the so-called conical
axis that runs parallel to the grooves of the grating (see also Fig.
\ref{fig:con_dif} and the explanation of this figure in the text). The
diffracted XUV beam is sent through slit S1 mounted in the focal plane of
the second toroidal mirror T2. The pair of toroidal mirrors T3+T4 produces
an image of the XUV beam selected by the slit in the interaction region of
the endstation that is connected to the monochromator beamline. The second
grating G2 is rotated by the same angle as grating G1 and compensates for
the temporal stretching induced by the first grating.  In our implementation
the grating pairs can be exchanged in place to optimize diffraction
efficiency. Three grating sets are used with the blaze angles of each
pair optimized for a different photon energy range. The parameters of the
three grating sets and of the toroidal mirrors are listed in Table
\ref{tab:optics}.

The transmission efficiency is an important factor for the design of the
two-grating monochromator, because diffraction and reflection efficiencies
in the XUV photon energy range are generally low even at grazing incidence.
To overcome this difficulty the diffraction gratings in the present layout
are used in an off-plane (also called conical diffraction) mount. The ray
geometry for the off-plane grating mount is shown in Fig. \ref{fig:con_dif}.
The grating is mounted with its grooves parallel to the long side of the
grating substrate, i.e. perpendicular to the traditional (in-plane)
diffraction configuration. The resolving power in this case is smaller than
in the traditional configuration using the same grating and beam size,
because less grooves are illuminated at the grazing incidence. For this
reason the off-plane mount is rarely used in situations, where bright
sources are available and efficiency can be sacrificed in favor of spectral
resolution. However in femtosecond time-resolved experiments the spectral
bandwidth required to support short pulses puts an upper limit on the energy
resolution that can be demanded in experiments, while the low yield of HHG
sources makes diffraction efficiency a more limiting factor. Here the
off-plane mount offers the advantage of very high reflectivity compared to
the traditional in-plane grating mount. It was shown that for a blazed
grating in the off-plane mount the diffraction efficiency in the direction
correspoinding to the blaze angle may reach the limit given by the
reflectivity of the coating (an Al coating in this case)
\cite{Werner1977,Neviere1978,Poletto2006,Pascolini2006}.

\begin{table}[htpb!]
\caption{Parameters of the optical elements of the TDCM.}
\label{tab:optics}
	\centering
%	\begin{tabular}{|c|c|c|c|c|c|c|}
%	\hline
%	   Optical Element &  
%           F\footnote{focal length, cm} &  
%           $\sigma$\footnote{groove density, grooves per mm} &
% 	   $\beta$\footnote{blaze angle} &  
%           $\gamma$\footnote{incident angle} &  
%           L\footnote{length, mm} &  
%           H\footnote{height, mm} \\
%	\hline
%	Toroidal mirror & 80  & - & - & 4\degree & 90 & 15 \\
% 	Grating I   & - & 150 & 3.4\degree & 5\degree & 70 & 15 \\
%	Grating II  & - & 300 & 4.3\degree & 5\degree & 70 & 15 \\
%	Grating III & - & 600 & 7.0\degree & 5\degree & 70 & 15 \\
%	\hline
%	\end{tabular} 
	\begin{tabular}{l|cccc}
	   & Toroidal & Grating I & Grating II & Grating III \\
	  \hline
	  focal length, cm & 80 & - & - & - \\
	  density, gr/mm & - & 150 & 300 & 600 \\
	  blaze angle & - & 3.4\degree & 4.3\degree & 7.0\degree \\
	  incident angle & 4\degree & 5\degree & 5\degree & 5\degree \\
	  energy range, eV & 3-50 & 14-30 & 20-38 & 32-50 \\
	  peak energy, eV & - & 18 & 28 & 35 \\
	  length, mm & 90 & 70 & 70 & 70 \\
	  height, mm & 15 & 15 & 15 & 15 \\
	\end{tabular}
\end{table}

In the off-plane geometry, diffraction of a ray incident on the grating
at a grazing angle $\gamma$ forms a cone with an opening angle of
$2\gamma$.
%The rays diffracted from a grating mounted in the off-plane geometry form a
%cone with an opening angle defined by the incidence angle $\gamma$. 
The grating diffraction equation in the off-plane case reads:
\begin{equation}
\sin{\gamma}(\sin{\theta}+\sin{\theta'})=m\lambda\sigma,
\label{eq:con}
\end{equation}
where 
%$\gamma$ is the incidence angle on the grating (the half-angle of the
%cone), 
  $\theta$ is the angle between projections of the incident ray and the
grating normal on a plane perpendicular to the grooves, $\theta'$ is the
angle between projections of the diffracted ray and the grating normal on
the same plane, $m$ is the diffraction order, $\lambda$ is the wavelength of
the diffracted light and $\sigma$ is the groove density. In the operation of
the monochromator the grating is rotated around an axis parallel to the
grooves and lying in the surface plane of the grating in such a way that the
diffracted ray corresponding to the wavelength of choice propagates at the
same angle as the incident ray (i. e. $\theta'=\theta$). For this particular
wavelength the diffraction equation simplifies to
  \begin{equation}
  2\sin{\gamma}\sin{\theta}=m\lambda\sigma.
  \label{eq:con2}
  \end{equation}
The angle $\theta$ in this equation corresponds to the rotation angle of the
diffraction grating, which is used to select the XUV wavelength $\lambda$.

Expected performance of the monochromator is estimated by numerical
ray-tracing of the optical setup using parameters of the optical elements
given in Table \ref{tab:optics}. The source size is estimated to be
60~$\mu$m for the XUV generation layout used in the present experiments. The
slit width is 100~$\mu$m. With these parameters the ray-tracing calculation
delivers estimates of the FWHM bandwidth of the XUV light passing through
the slit for a given XUV photon energy and the optical path diffrences
before and after the compensation in the second stage of the monochromator.
For the three gratings listed in Table \ref{tab:optics} the spectral
resolution in their corresponding energy ranges varies smoothly from 0.15 to
0.5~eV. The optical path difference after the first stage is approximately
100~fs for all gratings and is reduced to 2-4~fs by the compensation in the
second stage.

\section{Mechanical design and alignment}

%Even 0.2~mm of air absorbs
%more than 98\% of the XUV photons in the 10-50~eV energy range \cite{CXRO}.
%XUV radiation can only propagate in vacuum. 
%Therefore it is necessary to keep the 

In experiments with XUV light, the XUV source, the optical elements and 
the experimental target are located in a vacuum environment, which 
requires special care when constructing the optical mounts. Additionally, 
we choose to guide a second laser beam used for pump-probe experiments 
also in vacuum in close proximity to the XUV beam path to ensure 
long-term stability of the setup. All this puts stringent constraints on 
the geometry of the individual optical elements, the mechanical design 
and the alignment strategy of the TDCM. The most important elements in 
our implementation of the monochromator are the grating holders. In 
experiments, the incidence angles on both gratings are tuned precisely to 
select the ray with the desired photon energy in the first stage, to 
compensate the induced stretch in the second stage and to maintain the 
alignment of the ray with respect to the interaction region of the 
experimental endstation. Therefore stability of the grating holders, 
their alignment and reproducibility of rotation have a major impact on 
the overall performance of the TDCM, in particular, on keeping the 
position of the XUV focus fixed while scanning the XUV photon energy. 
This section describes the implementation and alignment of the the 
optical elements in our setup. The first subsection describes the design 
and pre-alignment of the grating holders. The complete beamline is 
described in the second subsection. The third and fourth subsections 
provide descriptions of the alignment of the main XUV arm and the second 
IR arm required for femtosecond pump-probe experiments.

\subsection*{Design and pre-alignment of the grating holders}

The correct operation of the monochromator requires that the XUV beam 
always follows the same optical path in the monochromator independent of 
the photon energy selected by the gratings. The path should also remain 
the same when switching between gratings with different groove densities. 
If the pointing of the XUV beam changed, it would lead to a displacement 
of the XUV beam in the interaction region and loss of the spatial overlap 
of the laser beams in the pump-probe experiment. Moreover, substantial 
changes in pointing of the XUV beam may cause aberrations in the imaging 
by the toroidal mirrors or lead to stretching of the pulse in the time 
domain. Parallel displacement of the XUV beam from the optical axis, on 
the contrary, is not very crucial as long as the geometrical aberrations 
for the off-axis rays remain small.

\begin{figure}
\centering
\includegraphics[width=8.5cm]{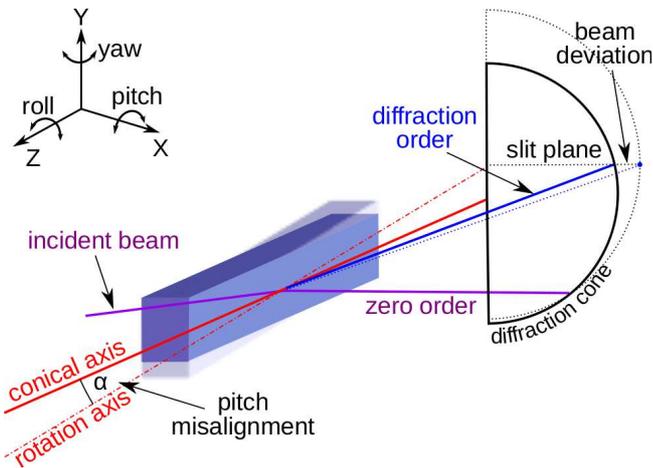}
\caption{\label{fig:grating_mis}
  Misalignment of the grating pitch. The conical axis of the grating is 
tilted with respect to the rotation axis by an angle $\alpha$. The new 
diffraction cone (solid line) has a different axis and a different 
opening angle as compared to the diffraction cone of the properly aligned 
grating (dotted line). The misalignment leads to a deviation of the 
diffracted beam with respect to the optical path of the TDCM (dotted 
line).
  }
\end{figure}

For the pointing of the XUV beam to remain constant for all wavelengths 
the rotation axis must coincide with the axis of the diffraction cone for 
the selected grating (see Fig. \ref{fig:con_dif}). If this is the case, 
then rotation of the grating moves diffracted rays on this cone. 
Consequently for every ray(wavelength) there is a rotation angle $\theta$, 
which sends that ray along the optical path of the monochromator (the beam 
path all elements are aligned to). On the other hand, if the rotation 
axis is misaligned with respect to the axis of the diffraction cone, the 
rotation changes the axis and the opening angle of the diffraction cone. 
This leads to a wavelength-dependent deviation of the diffracted beam 
from the optical path of the monochromator, i.e. the XUV beam pointing 
depends on the chosen wavelength. This effect is sketched in Fig. 
\ref{fig:grating_mis} for the case of a misalignment of the grating 
pitch. Similar effect is observed for a misalignment of the yaw.

It follows that the degrees of freedom provided by the grating holder 
should be used to align the grating grooves along the rotation axis of 
the grating holder. The displacement along the grating normal (X in Fig. 
\ref{fig:con_dif}) in this case controls the position of the grating surface 
with respect to the rotation axis and the pitch and yaw angles control 
the angle between the grating grooves and the rotation axis. We can show 
that for small grating rotation angles $\theta$ the grating pitch is the 
most critical parameter, while effects of a misalignment in the yaw are 
an order of magnitude smaller. For a pitch misalignment (shown in Fig. 
\ref{fig:grating_mis}) the change in the opening angle of the 
diffraction cone upon rotation of the grating can be calculated using 
the following expression:
 \begin{equation} 
 \frac{\cos \gamma'' - \cos \gamma'}{\sin \gamma'' + \sin \gamma'} = 
 \tan \alpha \sin \theta, 
 \end{equation} 
 where $\alpha$ is the pitch misalignment angle between the conical axis 
and the rotation axis, $\theta$ is the grating roll 
and $\gamma', \gamma''$ are half-angles of the diffraction cone before and 
after grating rotation, respectively. In the small angle approximation this 
expression simplifies to 
\begin{equation} 
\gamma'-\gamma'' = 2 \alpha \theta. 
\end{equation} 
  Typical spot sizes of the XUV and IR beams in pump-probe experiments 
are 100~$\mu$m and therefore the displacements of the XUV beam in the focus 
have to be limited to 10-20~$\mu$m. This is equivalent to XUV 
pointing deviations of less than 25 $\mu$rad. According to the equation above,
for a diffraction angle $\theta=100$~mrad the pitch misalignment angle $\alpha$
should be less than 125~$\mu$rad. 
%As mentioned above the condition on 
%the yaw misalignment is less stringent.

\begin{figure}
\centering
\includegraphics[width=8.5cm]{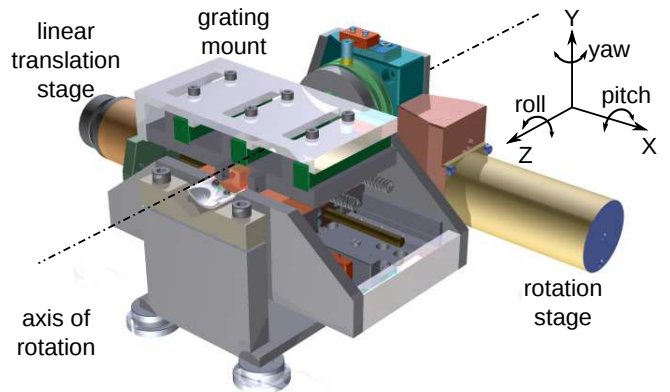}
\caption{\label{fig:SetupGra}
Assembly drawing of the grating holder. The grating mount housing three
gratings with different groove densities is attached to a linear
translation stage, which allows exchange of the gratings. The linear stage
is mounted on a rotation stage controlling the grating roll angle and thus the
selected XUV photon energy. The coordinate system used in the text is
defined in the upper right corner of the plot.
}
\end{figure}

Precise alignment of the pitch and yaw angles and the position in the 
direction X (for all six gratings in the two stages of the TDCM) are thus 
required for correct operation over the full wavelength range. Each holder 
houses three gratings and therefore should in principle provide tunability 
for $3 \times 3=9$ degrees of freedom. To reduce the mechanical complexity 
of the holders and ease the alignment, we decided to mount the three 
gratings of each monochromator stage in one common fixed mount, which is 
precisely machined to ensure that the grating surfaces and their grooves are 
parallel to each other. The design of the grating holder is shown in Fig. 
\ref{fig:SetupGra}. 
%Each holder houses in the fixed mount three gratings 
%with different groove densities.
 The gratings are exchanged by linearly translating the mount by means of a 
vacuum-compatible linear translation stage (Micos, model LS-65 UHV). The 
rotations of the grating are performed by a vacuum-compatible rotation 
stage (Micos, model DT-65N UHV), which holds the linear stage that carries 
the gratings. The parallelism of the gratings is checked by examining the 
reflection of an alignment laser (HeNe laser) from each of the three 
gratings. The roll angle (see Fig. \ref{fig:SetupGra}) showed slight 
deviations, which could be corrected by calibration of the rotation stage. 

\begin{figure*}[!t]
\centering
\includegraphics[width=\textwidth]{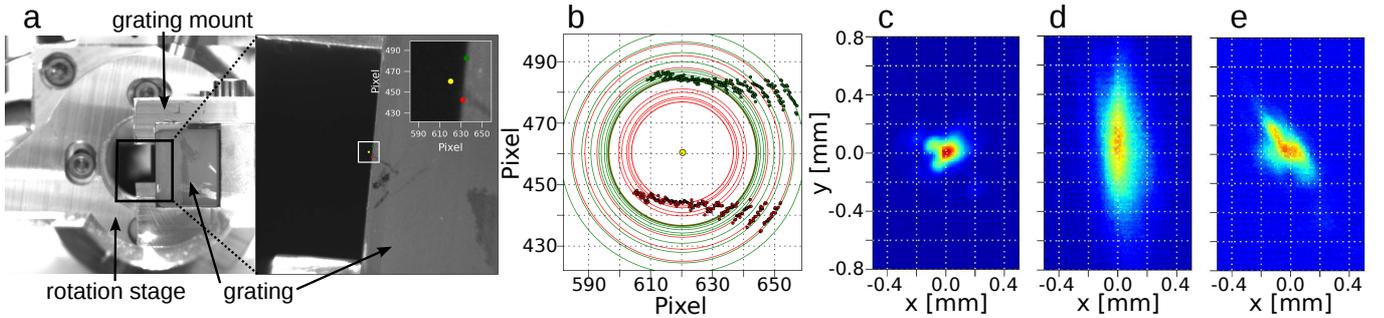}
\caption{\label{fig:SetupGra_align}
Alignment of the grating holder. a) High-resolution close-up images of the grating
edge with two distinct points (red and green) marked. b) Trajectories followed
by two points marked in a), when the grating is rotated around the roll axis
and/or displaced along the direction X. The concentric circles fitted to the
trajectories help to establish the position of the rotation axis (yellow point). 
Alignment using the 400~nm light.
c) The profile of the 400~nm beam for the grating
turned to zero order. The image distortions are caused by diffraction on
grating edges. d) The beam stretched by diffraction from
the first grating. The beam is centered at zero indicating correct alignment
of the grooves (see text). e) The beam stretched by the first grating and 
recompressed by the second grating.
}
\end{figure*}

The holders with fixed grating mounts reduce the number of required 
degrees of freedom to four: the displacement in the X direction and the 
roll, pitch and yaw angles for all gratings simultaneously. The roll 
angle is motorized and is used to tune the monochromator to the desired 
wavelength. Therefore it does not need alignment, but instead requires 
wavelength calibration.

Alignment of the grating displacement along X and of the grating yaw is
performed by imaging the edge of the grating by a CCD camera through a zoom
lens with a large magnification, as shown in Fig. 
\ref{fig:SetupGra_align}a. The focal plane of the camera is approximately
perpendicular to the rotation axis of the holder and therefore the edge
remains in focus when the grating is rotated. For a set of linear stage
positions (X) and grating rotation angles $\theta$ positions of two
recognizable marks on the grating edge (small scratches, labeled by red and
green dots in Fig. \ref{fig:SetupGra_align}a) are traced. When the grating
is rotated, these marks follow circular trajectories shown in Fig. 
\ref{fig:SetupGra_align}b for several linear stage positions. The common center
of these circles corresponds to the position of the rotation axis in the
focal plane of the CCD camera. The position of the rotation axis is thus
extracted with a precision of 5~$\mu$m, estimated by repeating the
measurement several times for slightly diffrerent CCD camera alignments. 
This procedure is performed for both edges of the same grating and gives the
optimal linear stage position X for each edge. The difference of these
optimal positions is used to calculate the residual yaw of the grating,
which is then corrected by tilting the grating mount. The estimated
precision of 5~$\mu$m corresponds to an angular precision of 70~$\mu$rad,
which is less than the limiting misalignment angle given above.

As mentioned above, the most crucial alignment parameter is the pitch. 
The high groove densities of the gratings prevent use 
of optical imaging methods for groove alignment. To overcome this difficulty 
the grating holders are designed to reach rotation angles of more than 
25{\degree}, which is sufficient to observe the first order diffraction of a 
400~nm beam from the grating with the lowest groove density of 150 grooves per 
mm. Therefore a 400~nm beam, which is easy to generate with a femtosecond 
laser, can be used to align the grating pitch with the required precision of 
less than 125~$\mu$rad.

In this alignment procedure the 400~nm beam is first reflected from the 
grating rotated to the zero order. The position of the beam is recorded on 
a high-resolution beam profiler. The grating is then rotated to the angle 
corresponding to the first order diffraction of the 400~nm beam 
($\theta=20.1$\degree). The horizontal displacement of the diffracted beam 
from the position of the zero-order beam is due to the pitch 
misalignment of the grating and therefore can be used to adjust the 
grating.

The 400~nm beam profiles recorded in this alignment procedure are shown 
in Fig. \ref{fig:SetupGra_align}c and d. As required, the beams are 
located precisely at the same spot on the beam profiler, but the 
diffracted beam (Fig. \ref{fig:SetupGra_align}d) is visibly elongated 
due to spectral dispersion on the first grating. The alignment procedure 
is repeated for the second grating holder.  Fig. 
\ref{fig:SetupGra_align}e shows the beam profile after installing both 
grating holders in the 400~nm beam path and rotating both gratings to the 
first diffraction order. The beam is again positioned precisely at zero 
and is vertically confined, because the second grating compensates the 
spectral dispersion induced by the first grating. The increased 
distortions of the beam profile are caused by diffraction on edges of 
both gratings. In the case of the XUV beam these effects are 
significantly reduced compared to those observed for the 400~nm beam, 
because of smaller XUV beam size and shorter wavelength. Thus aligned the 
grating holders can be moved and placed at the respective positions in 
the XUV monochromator beamline.

\subsection*{Implementation of the monochromator beamline}

\begin{figure*}[!t]
\centering
\includegraphics[width=\textwidth]{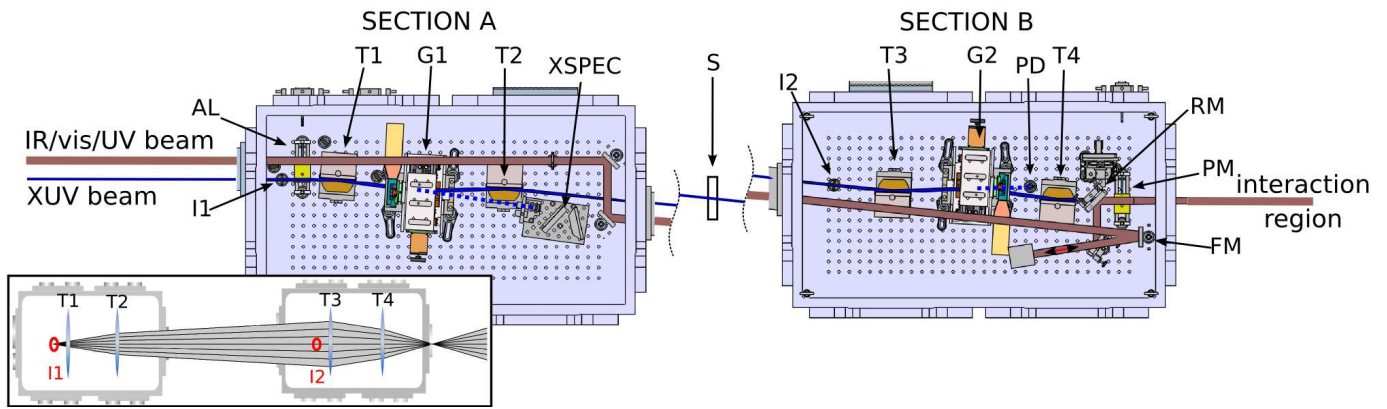}
\caption{\label{fig:full}
Schematic drawing of the TDCM beamline:
AL - aluminium filter;
T1,T2,T3,T4 - toroidal mirrors;
G1,G2 - gratings;
I1, I2 - XUV alignment irises;
S - slit;
XSPEC - XUV spectrometer;
PD - XUV photodiode;
RM - recombination mirror;
FM - focusing mirror;
PM - pick-off mirror.
The inset shows the ray-tracing of the TDCM alignment used to choose the 
position of the iris I2 (see text for details).
}
\end{figure*}

The complete design of the TDCM beamline is shown in Fig. \ref{fig:full}. A 
commercial state-of-the-art laser system (Aurora, Amplitude Technologies) 
provides ultrashort pulses with 795~nm central wavelength and durations of 
25~fs at 1kHz repetition rate. A beam with a pulse energy of about 1~mJ is 
focused in a gas cell filled with a noble gas (typically argon) to produce 
the XUV light via high-order harmonic generation (HHG)\cite{Pfeifer2006}. 
The XUV beam propagates towards the first section of the monochromator 
(section A), where the residual IR beam is blocked by a 100~nm thick Al 
filter (Lebow), which is transparent for XUV photons in the range of 20 to 
72~eV. The filter helps to reduce the scattered IR light 
and protects the first toroidal mirror T1 from Al deposits, which propagate 
along the laser beam from the HHG cell. The filtered XUV beam is collimated 
by the first toroidal mirror (T1) positioned 80~cm away from the focus in 
the HHG cell. The collimated beam impinges on the first diffraction grating 
G1 at an angle of 5$^\circ$ and is then refocused in the plane of the 
monochromator slit (S1) by the second toroidal mirror (T2).

In the focal plane the rays diffracted from the grating G1 are located on a
cone as shown in Fig. \ref{fig:con_dif}. A horizontal slit positioned
perpendicular to the cone selects the ray corresponding to the desired
photon energy while blocking all other rays. The energy resolution of the
monochromator can be adjusted by selecting one of the four slits with fixed
widths (50, 100, 200 and 650 $\mu$m) mounted on a motorized manipulator.
The expected photon energy resolution for the 100~$\mu$m slit ranges from
0.15 to 0.5~eV according to ray-tracing estimates given above. This value
increases approximately linear with width for wider slits, but only marginal 
improvement is expected for the 50~$\mu$m slit, beacause of geometrical
aberrations induced by toridal mirrors.

The XUV beam passing through the slit is collimated by the third toroidal 
mirror (T3) and impinges on the second grating G2. The grating G2 is always 
rotated by the same angle as grating G1 but in the opposite sense of 
rotation. This ensures that the temporal stretch induced by diffraction on 
the first grating is compensated by the second grating. The beam diffracted 
from G2 is focused by the last toroidal mirror T4 into the interaction 
region. On its way the beam passes through a hole drilled at 45$^\circ$ in 
a 2-inch recombination mirror (RM).

The are two auxiliary tools for characterization of the XUV beam, which are 
implemented in our setup. As can be seen in Fig. \ref{fig:full} the linear 
translation stage of the grating holder in section A can be used to 
displace the grating from the path of the XUV beam, which allows the beam 
to reach the entrance slit of a compact XUV spectrometer (XSPEC) \cite{Kornilov2010}. 
The spectrometer is used to monitor the full HHG spectrum at the entrance 
of the monochromator. Similarly, in section B, the XUV beam passing by 
the grating G2 can reach a calibrated XUV photodiode (AXUV-100, 
International Radiation Detectors), which is used to monitor the absolute 
XUV pulse energy passing through the slit of the monochromator. The 
measured value should be corrected for the diffraction efficiency of 
grating G2 and the reflectivity of the last toroidal mirror T4 to obtain 
the XUV pulse energy at the interaction region.

In the present experiments a velocity map imaging spectrometer (VMI) 
\cite{Eppink1997} is connected at the end of the monochromator beamline and 
used for characterization of the beamline, as well as execution of first 
experiments \cite{Eckstein2015}. The VMI spectrometer can record kinetic 
energy and angular distributions of both electrons and ions resulting from 
photoionization by XUV pulses. The target in the present case is a gas jet 
originating from a 10~$\mu$m orifice in the repeller plate of the 
spectrometer \cite{Ghafur2009}.

\subsection*{Alignment}

Initial alignment of the TDCM components at their design positions is 
performed using a HeNe alignment laser. A calibrated 100~$\mu$m pinhole is 
placed at the position of the HHG cell to define the source size. The beam 
is imaged by a beam profiler after each optical element to ensure proper 
alignment. Toroidal mirrors are mounted in five-axis vacuum-compatible 
mirror holders (Alca) and their positions are tuned to minimize imaging 
aberrations as observed on the beam profiler. The pre-aligned grating 
holders are placed in their positions ensuring that the incidence angles 
are equal to the design value of 5\degree.

Operation of the TDCM requires not only careful alignment of all optical
components, but also precise alignment of the driving femtosecond IR beam
(and thus the XUV beam) to the optical axis of the setup. This operation
should be performed daily and without breaking the vacuum. For this purpose
we employ two irises, labeled I1 and I2 in Fig. \ref{fig:full}, which are
imaged by CCD cameras via vacuum viewports. The first iris is positioned at
the entrance to section A of the monochromator before the first toroidal
mirror T1. The position of the second iris I2 is chosen such as to increase the
precision of alignment. For this a simple 2D ray tracing is performed
with the toroidal mirrors replaced by ideal lenses and the gratings removed for
simplicity. The results are shown in the inset of Fig. \ref{fig:full}. 
The rays start at the position of the first iris I1
assuming perfect alignment at that iris and follow slightly diverging
paths. The second iris I2 is positioned in front of mirror T3,
where the spread of the rays is largest, thus maximizing the sensitivity to
the deviations from the optical axis. This procedure ensures alignment of
the XUV beam with a precision better than 80~$\mu$rad, sufficient for correct
operation of the TDCM.

%\begin{figure}
%\centering
%\includegraphics[width=\columnwidth]{figs/fig_alignment.png}
%\caption{\label{fig:alignment}
%Ray-tracing simulation of the TDCM alignment. The beam is considered to be
%perfectly aligned to the first iris I1. The position of the second iris I2
%is chosen at the location of largest divergence of the rays. The toroidal
%mirrors are replaced by lenses and the gratings are removed for simplicity.
%}
%\end{figure}

\subsection*{The IR/vis/UV arm for the pump-probe experiments}

The main purpose of the present TDCM setup is investigation of 
femtosecond molecular dynamics using pump-probe spectroscopy with XUV 
pulses. For this purpose a second laser arm is constructed, which is 
designed to transmit laser pulses in the near-IR, visible or UV ranges. 
For stability and laser safety the beam in this arm is guided 
through the same vacuum system as the XUV beam, as shown in Fig. 
\ref{fig:full}. The beam enters the vacuum system through a viewport 
located in the vicinity of the entrance window of the HHG driving laser. 
It is guided by two 2-inch mirrors and is incident on a focusing mirror 
(FM) located at the far end of section B. The converging beam is then 
recombined with the XUV beam by the recombination mirror RM. For precise 
control of the focus position, the divergence of the second beam is adjusted 
by a two-mirror telescope before the beam enters the vacuum system.

%\begin{figure}
%\centering
%\includegraphics[width=\columnwidth]{figs/cross_correlation_1_cut.png}
%\caption{\label{fig:H2}
%\note{H2 signal, data and fit}
%}
%\end{figure}

To establish the spatial and temporal overlap of the two pulses, a pick-off
mirror (PM) can be positioned in the beam after the recombination mirror by
means of a manual manipulator. The XUV beam cannot pass through the glass 
viewport, but when the aluminium filter is moved out of the beam and
gratings G1 and G2 are tuned to zero order, the driving IR beam propagates
through the monochromator and can be used for pre-alignment of the
pump-probe overlap. Fine alignment of the spatial and temporal overlap
uses the process of bond softening in H$_2^+$ molecules \cite{Johnsson2010}. 

%In this experiment the H$_2$
%molecules are ionized by the XUV pulse and subsequently dissociated by a
%moderately strong 800~nm pulse. The yield of H$^+$ ions as a function of
%pump-probe delay is recorded by the VMI spectrometer and shown in Fig. 
%\ref{fig:H2}. 

\section{Characterization}

Three main parameters characterize the performance of the XUV monochromator:
the spectral resolution, the temporal resolution and the total transmission
of the monochromator beamline. This section describes results of performance
tests for these three parameters. All of them depend on the XUV photon
energy and therefore the transmission and the spectral resolution of the
setup are characterized for several photon energies in the tuning range of the
monochromator. The temporal characterization, however, is quite
complex and only performed for a photon energy of 29.6~eV, corresponding to
the 19th harmonic. These experiments demonstrate the compensation scheme of
the TDCM and yield a duration of the XUV pulse of $12\pm3$~fs in good
agreement with the previous experiments, which demonstrated the duration of
$13\pm0.5$~fs for the 19th harmnonic generated at similar experimental
conditions \cite{Poletto2009a}. Further experiments outside of the scope of
this work will be performed to test practical limits of the temporal
resolution, which are estimated in the Optical design section.

\subsection*{Transmission of the monochromator}

The conical diffraction principle used in the present design of the TDCM 
allows for very high transmission of the XUV radiation even though six 
optical elements are used in the optical layout. The transmission is 
further enhanced by using three sets of gratings each optimized for a 
specific photon energy range. The XUV transmission for the present setup is 
characterized using two calibrated XUV photodiodes (AXUV-100, International 
Radiation Detectors). Since both sections of the monochromator are 
assembled from identical elements, only section B is characterized and the 
resulting transmission coefficient is squared to obtain transmission of the 
complete monochromator \cite{Kubin2013}.

\begin{figure}
\centering
\includegraphics[width=8.5cm]{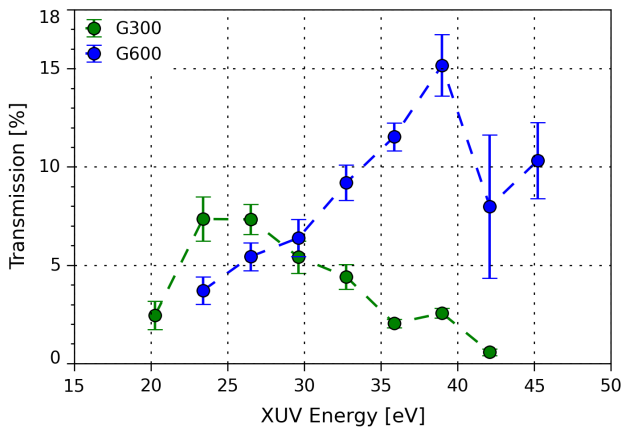}
\caption{\label{fig:transmission}
Transmission coefficients of the full TDCM beamline for the 300 gr/mm
(green) and 600 gr/mm (blue) gratings as a function of the XUV photon energy
\cite{Note1}. The optimal photon energy ranges for the gratings are listed 
in Table \ref{tab:optics}.
}
\end{figure}

The first photodiode is mounted on a mechanical manipulator and located after the 
monochromator slit before the toroidal mirror T3. It measures the total XUV flux 
entering section B.  The second photodiode is mounted after the last toroidal 
mirror T4 and measures the flux delivered to the interaction region. The 
measurement is performed for two of the three sets of gratings
  \footnote{The transmission measurement for the photon energy range of the 150 
  gr/mm grating could not be carried out because the Al filter, which is
  required to protect the photodiodes from the residual IR light, has very
  low transmission in the energy range of this grating. This poses no problems 
  for using the grating in experiments, which are not sensitive to a stray 
  IR light.}
 by tuning the 
gratings from one HHG peak to the other and measuring the incoming and outgoing 
photon fluxes. The ratio of the two fluxes gives the transmission coefficient for 
section B.
% and the square of it represents the transmission of the complete 
%TDCM. 
%For this measurement the size of the slit is chosen to be 200~$\mu$m, 
%sufficiently wide to let the full spectrum of an individual harmonic pass 
%through. 
  The results of the transmission measurements are shown in Fig. 
\ref{fig:transmission}. They demonstrate that the transmission is above 3\%
in the range of the measurement and reaches a value of 15\% for the harmonic, 
which has its diffraction angle close to the blaze angle of the 600 gr/mm grating
\cite{Note1}.

\subsection*{Spectral characterization}

\begin{figure}
\centering
\includegraphics[width=8.5cm]{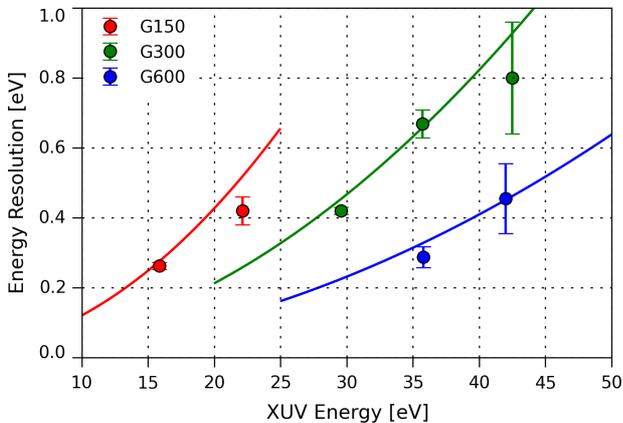}
\caption{\label{fig:eng_res}
Energy resolution of the TDCM employing the 100~$\mu$m slit. The round
symbols correspond to the gratings with groove densities of 150 gr/mm (red),
300 gr/mm (green) and 600 gr/mm (blue). The lines represent estimates of the
resolution as described in the text.
}
\end{figure}

%The bandwidth of the XUV light passing through the monochromator depends on
%the groove density of the gratings and the width of the slit used for
%selecting individual harmonics. 

The spectral resolution of the monochromator (defined as the full width at half 
maximum (FWHM) of the spectrum) is limited by the resolving power of the gratings, 
the XUV source size and aberrations of the focusing optics. These effects put 
practical limits on the width of the slit used to filter the desired XUV 
wavelength. The experimental measurement of the spectral resolution employs 
XUV photoionization of helium and xenon atoms in the interaction region of the 
velocity map imaging spectrometer. The kinetic energy of the photoelectrons is 
equal to the difference between the XUV photon energy and the ionization 
potential (IP) of the target atom. Therefore, the width of the photoionization 
peak detected in the VMI spectrometer is directly related to the bandwidth of 
the XUV light passing through the slit of the monochromator beamline. The VMI 
spectrometer additionally broadens the photoelectron peaks by about 2\% of the 
kinetic energy of the electron. The target gas is chosen such as to minimize 
the kinetic energy of the photoelectrons and thus reduce the effect of the VMI 
energy resolution. For the high photon energies (harmonics 19 to 27) He atoms 
with an ionization potential (IP) of 24.59~eV are used.  For lower harmonics 
(11th and 15th), which have an insufficient photon energy to ionize helium, we 
employ Xe atoms with an IP of 12.13~eV.  The ground state of Xe ions has a 
spin-orbit splitting of 1.31~eV.  This splitting is sufficiently large and can 
be fully resolved in the VMI spectra. Therefore it does not affect the 
monochromator resolution measurements.

In the experiments, first the bandwidth of the 19th harmonic is analyzed using 
the 300 grooves/mm gratings and four different slits with widths of 650, 200, 
100 and 50~$\mu$m. Analysis of the photoelectron spectra shows that the 
bandwidth of the XUV light passing through the beamline is reduced when going 
from 650~$\mu$m to 200~$\mu$m slit and from 200~$\mu$m to 100~$\mu$m slit, but 
the bandwidth is practically identical for the 100~$\mu$m and 50~$\mu$m slits. 
We therefore conclude, that 100~$\mu$m is the practical limit for the slit 
width, at least for the case of the 19th harmonic. Next, for each grating pair 
two or three harmonics are selected from the XUV spectrum using the 100~$\mu$m 
slit and the width of the photoelectron peak is recorded in the VMI. For each 
combination of harmonic and grating, photoelecrton spectra are also 
recorded without the slit to determine the input XUV bandwidth. It is 
important to ensure that the input bandwidth is substantially wider than the 
bandwidth selected by the 100~$\mu$m slit, since otherwise the spectral shape 
of the input radiation significantly influences the bandwidth of the output 
and invalidates the resolution measurement.

The results of the experimental series are presented in Fig. 
\ref{fig:eng_res} as filled circles with colors corresponding to the 
three grating sets. The figure also includes estimates of the spectral 
resolution, which are calculated by ray-tracing of the optical layout 
with the assumptions that the XUV source size is equal to 60~$\mu$m 
and that the XUV beam divergence is 1~mrad independent of the harmonic 
order (estimated from the focusing conditions of the driving IR beam 
\cite{Salieres1996}). The geometric aberrations, which limit spectral 
resolution by increasing the source image size at the position of the 
slit, are estimated at the alignment phase and amount to 90~$\mu$m 
(FWHM image size of an infinitely small source) independent of the XUV 
photon energy. Fig. \ref{fig:eng_res} demonstrates that the results of 
these estimates are in very good agreement with the measurements, 
which serves as additional confirmation of the proper alignment of the 
monochromator beamline. The optimal spectral resolution of the 
monochromator is less than 500~meV for photon energies below 45~eV and 
increases to 600~meV at the limiting photon energy of 50~eV. The plot 
in Fig. \ref{fig:eng_res} combined with the transmission data in Fig. 
\ref{fig:transmission} can serve as a guide for choosing the grating 
groove density and the slit width for the specific XUV photon energy.

% \note{Markus, IR divergence: 6.4+-0.8 mrad} 

\subsection*{Temporal characterization}

The temporal characterization of the monochromator is performed using the 
process of sideband generation \cite{Kroll1973,Maquet2007}.  In 
experiments of this kind, the atoms of a rare gas (in this case argon) are 
ionized by the XUV pulse in the presence of a moderately strong IR pulse.  The 
electric field of the IR pulse modifies the kinetic energy of the 
photoelectrons resulting in additional features (sidebands) in the 
photoelectron kinetic energy distribution.  The kinetic energies of the 
sidebands are shifted with respect to the main photoline by an integer number of 
IR photons, both to lower and higher energies.

\begin{figure}
\centering
\includegraphics[width=2.5in]{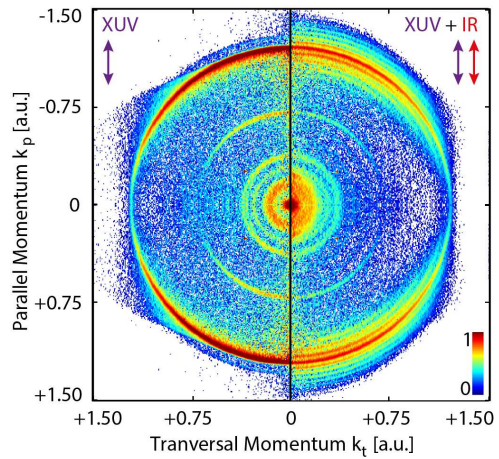}
\caption{\label{fig:MK_sidebands}
Left: momentum map of photoelectrons produced by ionization of argon atoms by
the 23rd harmonic (35.9~eV). The polarization of the XUV light is vertical.
Right: similar momentum map in presence of the strong IR field. Sidebands appear
as concentric rings in the vicinity of main photolines.
}
\end{figure}

Fig. \ref{fig:MK_sidebands} shows photoelectron VMI images recorded for 
ionization of Ar atoms by harmonic 23 (35.9~eV) without the IR field (left 
side) and in the presence of the IR pulse at zero time delay (right side).  
The image without the IR shows two distinct outer rings and three rings 
close to the center (at low kinetic energies).  The outermost ring 
corresponds to photoionization producing the Ar$^+$ (3s$^2$ 3p$^5$ $^2$P) 
ionic final state.  This state has a spin-orbit splitting of about 180~meV, 
which cannot be resolved in the present experiments.  The second ring 
corresponds to the Ar$^+$ (3s$^1$ 3p$^6$ $^2$S) ionic state.  The 
structures with low kinetic energies can be assigned to the formation of 
several electronically excited states of Ar$^+$. In the right half of 
Fig. \ref{fig:MK_sidebands} the image is strongly modified by the presence 
of an IR pulse with an energy of 130~$\mu$J. The main rings corresponding 
to single-photon ionization by the XUV pulse are weaker and many additional 
concentric rings (sidebands) appear with an energy spacing 
of the IR photon energy.
	
The temporal properties of both the XUV and IR pulses are investigated by 
scanning the time delay between the pulses.  The variation of the intensity 
of the IR field at the time of the arrival of the XUV pulse is reflected in 
the intensities of the sidebands.  For a sufficiently long IR pulse and high 
XUV photon energy the intensity of the n-th sideband for a fixed 
photoemission angle depends on the electric field of the IR pulse as 
follows \cite{Kroll1973,Maquet2007}:
  \begin{equation} 
  S_n(\vec k) \propto J_n^2 \left( x \right), 
  x = \frac{\vec E \cdot \vec k}{\omega^2},
  \end{equation} 
  where $\vec k$ is the momentum of the electron, $\vec E$ is the maximum 
electric field experienced by the electron, $\omega$ is the frequency of 
the IR field (all quantities are in atomic units) and $J_n(x)$ is the n-th 
order Bessel function. For sufficiently weak electric fields used in this 
experiment, the argument of the Bessel function is small and $J_n^2(x)$ is 
proportional to $x^{2n}$, i.e. to the IR pulse intensity.

\begin{figure}
\centering
\includegraphics[width=8.5cm]{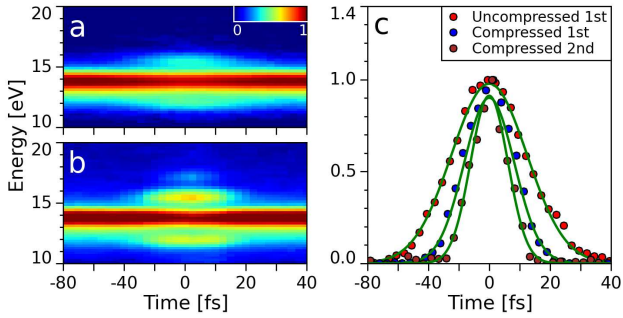}
\caption{\label{fig:sidebands}
Time- and energy- resolved sidebands generated in argon by the 19th harmonic
(29.6~eV). The sideband intensity is integrated over all emission angles.
a) Transient map for the fully compressed XUV beam. b) A similar
map for the uncompressed XUV beam (the second grating is at zero order). c) 
Temporal profiles of the first and second positive sidebands of the
compressed pulse and of the first positive sideband of the uncompressed
pulse. The profiles are closely represented by Gaussian distributions with
parameters described in the text.
}
\end{figure}

The maximum electric field experienced by the photoelectron varies with 
time delay according to the envelope of the the electric field of the 
pulse. The temporal profile of the n-th sideband can be calculated as a 
convolution of the XUV pulse profile with the IR pulse intensity profile 
raised to the power of $n$.  If both XUV and IR pulses have Gaussian shapes 
with a FWHM of $\Delta t_{XUV}$ and $\Delta t_{IR}$, the temporal profile of 
the first and second sidebands are given by Gaussian distributions with 
widths:
  \begin{eqnarray}
  \Delta t_1^2&=&\Delta t_{XUV}^2+\Delta t_{IR}^2 \nonumber \\ 
  \Delta t_2^2&=&\Delta t_{XUV}^2+\Delta t_{IR}^2/2.
  \label{eq:sig}
  \end{eqnarray}
 These expressions allow a calculation of both the XUV and IR pulse 
durations. Fig. \ref{fig:sidebands}a shows a false color map of the 
photoelectron kinetic energy distributions as a function of XUV-IR 
time delay measured with 300~gr/mm gratings and a 100~$\mu$m slit. The 
temporal profiles of the two positive sidebands (those shifted to 
higher energies with respect to the photoline) are plotted in Fig. 
\ref{fig:sidebands}c along with Gaussian fit functions. The extracted 
full widths at half maximum (FWHM) for the two sidebands are $\Delta 
t_1=38.8\pm0.4$~fs and $\Delta t_2=28.8\pm0.6$~fs. Using the 
expressions in Eq. \ref{eq:sig} values of $\Delta t_{IR}=37\pm1$~fs 
and $\Delta t_{XUV}=12\pm3$~fs are extracted for the FWHM of the IR 
and XUV pulses, respectively. The width of the XUV pulse is remarkably 
close to the value of $13\pm0.5$~fs demonstrated previously using a 
similar time delay compensating monochromator \cite{Poletto2009a}. 
This measurement indicates, that our IR pulse is somewhat stretched 
due to slightly unequal amounts of dispersive material in the XUV and 
IR beam paths.

To quantify the compensation effect induced by the second stage of the 
TDCM, temporal profiles of the sidebands are measured with the second 
grating G2 positioned at zero angle (no compensation).  The false color 
map of delay-dependent photoelectron kinetic energy distributions is 
shown in Fig. \ref{fig:sidebands}b.  The temporal profile of the first 
order sideband is plotted in Fig. \ref{fig:sidebands}c and is visibly 
wider than the profiles obtained in the experiment with compensation.  
The measured FWHM of the sideband is $60\pm1$~fs, which results in a 
temporal duration of the stretched XUV pulse of $47\pm1$~fs, i.e.  the 
XUV pulse is stretched by a factor of 3.1 upon diffraction off grating 
G1.  This stretching corresponds to illumination of 340 grooves or a spot 
size of about 1.1~mm on the 300 grooves/mm grating used in the present 
experiments. This number is compatible with the XUV spot size of 0.8~mm 
roughly estimated from the divergence of the driving IR beam 
\cite{Salieres1996}. 

The results of temporal characterization demonstrate, that the monochromator
preserves the XUV pulse duration within the accuracy of the
current measurement ($\pm3$~fs). According to the ray-tracing results
presented above the theoretical limit for the current design is on the order
of 2-4~fs depending on the XUV photon energy. This limit will be practically
tested in future experiments, which will employ sub-10~fs IR pulses for
generation and characterization of the temporal resolution.

\section{Application to sideband generation in IR-assisted XUV ionization of
argon atoms}

\begin{figure}[!t]
\centering
\includegraphics[width=8.5cm]{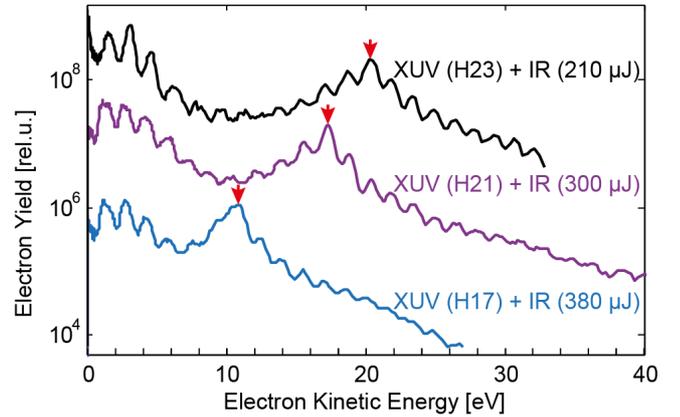}
\caption{\label{fig:HH_sidebands}
Ionization of argon atoms by harmonics 17, 21 and 23 in the presence of a
strong IR field (pulse energies are shown in the labels). The 
photoelectron distributions are integrated in the angular range of
$\pm15${\degree} from the polarization axis.
The positions of
the main photolines are marked by red arrows. The structures at low kinetic
energies are attributed to above-threshold ionization of argon atoms by the
IR pulses.
}
\end{figure}

The performance of the TDCM setup is demonstrated by investigating the
process of sidebands generation in IR-assisted XUV ionization of argon atoms
\cite{Kubin2013}. In the previous section this process was used for temporal
characterization employing relatively weak IR pulses, which results in only
a few sideband orders. When the energy of the IR pulse is increased, many more
sidebands can be observed. Fig.  \ref{fig:HH_sidebands} shows photoelectron
kinetic energy spectra for Ar atoms ionized by the 17th, 21st and 23rd harmonics
in the presence of an IR field.
Up to 13 sidebands can be seen accompanying the
main photoemission line corresponding to the Ar$^+$ (3s$^2$ 3p$^5$ $^2$P) final
state (labeled by an arrow), which shifts with the energy of the XUV photon.

\begin{figure}[!tb]
\centering
\includegraphics[width=6.5cm]{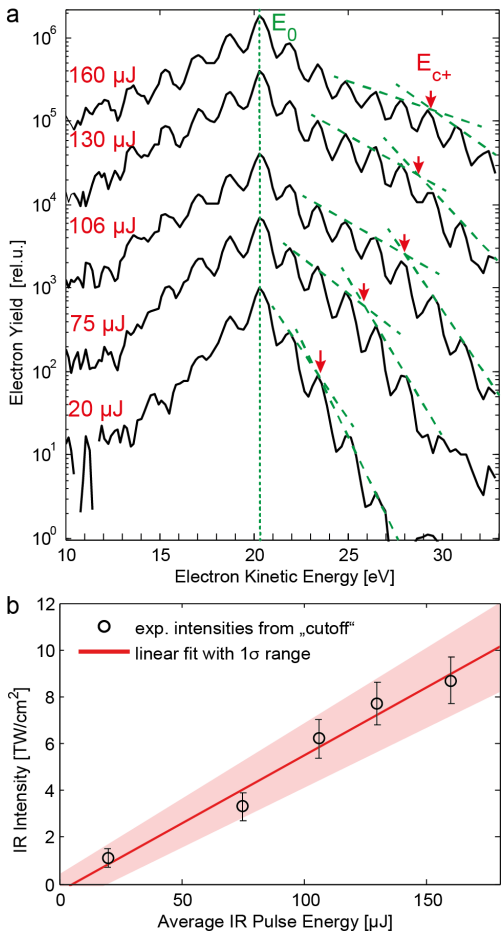}
\caption{\label{fig:IRpower}
Calibration of the intensity of the 800~nm IR pulse using a ''cut-off'' law
for the sideband generation process. a) Sideband intensity distributions
plotted on a logarithmic scale for five pulse energies from 20~$\mu$J to
160~$\mu$J. The photoelectron distributions are integrated in the angular 
range of $\pm10${\degree} from the polarization axis. The spectra are
shofted with respect to each other for better visibility.
The positions of the ''cut-off'' energy $E_{c+}$ for each curve
are labelled with arrows. b) IR intensity derived from the ''cut-off''
energies for the five pulse energies. Linear regression provides the scaling
of the IR intensity at the focus as a function of the pulse energy. The
non-zero origin of the regression line is due to a systematic offset in the
powermeter reading.
}
\end{figure}

The sideband intensity and their number depend on the electric field of 
the IR pulse at the moment of ionization by the XUV pulse 
\cite{Kroll1973,Maquet2007}.  Therefore sideband measurements can be used 
to determine the IR laser pulse intensity at the focus, knowledge of 
which is essential for design and implementation of most pump-probe 
experiments. Electron kinetic energy distributions for several different 
IR pulse energies are shown in Fig. \ref{fig:IRpower}a. Only few 
sidebands are generated at the lowest pulse energy of 20~$\mu$J. For 
higher pulse energies the number of sidebands substantially increases. 
Plotted on a logarithmic scale, the fall-off of the sideband intensity 
shows a characteristic piecewise linear behavior \cite{Radcliffe2012}. 
The slope of the fall-off changes approximately at the positions 
indicated by arrows. These energies, called ''cut-off'' energies, are 
related to the maximum kinetic energy, which an electron created by the 
XUV pulse can classically acquire through interaction with the IR pulse. 
For positive and negative sidebands they are expressed (in atomic units) 
as \cite{Maquet2007,Radcliffe2012}:
  \begin{equation}
  E_{c\pm} = E_0 \pm \sqrt{8 E_0 U_p} + 2U_p,
  \end{equation}
  where $E_0$ is the initial energy of the photoelectron given by the difference
between the XUV photon energy and the ionization potential and 
\begin{displaymath}
U_p[\mathrm{eV}] = 9.338 \times 10^{-8} \times (\lambda_{IR}[\mathrm{nm}])^2
\times I_{IR}[\mathrm{TW/cm}^2]
\end{displaymath}
is the ponderomotive potential of the IR field.

Thus the value of the cut-off energy can be used to calculate the 
average intensity of the IR field at the focus. This method is more reliable 
than estimations
based on the laser beam parameters such as pulse duration and size
of the focal spot. Fig. \ref{fig:IRpower}b shows the field intensity
calculations based on the ''cut-off'' law for five values of the pulse
energy.  As expected, the intensity is linearly proportional to the IR pulse
energy. The coefficient of proportionality of $67\pm6$~GW/cm$^2$/$\mu$J
gives an estimate of the beam FWHM at the focus of 140~$\mu$m
\cite{Kubin2013}.

\section{Conclusions}

In conclusion, we have implemented a time delay compensating XUV 
monochromator suitable for femtosecond pump-probe experiments with 
wavelength-selected XUV pulses. The beamline is designed to operate in 
the photon energy range from 3 to 50~eV. The XUV radiation is produced 
by means of high-order harmonic generation, which is demonstrated to 
deliver XUV pulses with durations of $12\pm3$~fs for the harmonic 19. 
This duration is 
compatible with the expected duration of the XUV pulse generated at the 
present exprimental conditions and is preserved by the monochromator 
beamline thanks to the two-grating design. The transmission of the 
beamline varies from 3\% to 15\% depending on the photon energy. The 
optimal spectral resolution of the beamline varies from 300~meV to 
500~meV FWHM for the photon energies below 45~eV and increases 
to 600~meV at the limiting photon energy of 50~eV.

We analyze the most sensitive parameters for TDCM alignment, present the
ray-tracing calculations and practical implementation of the alignment
procedure, which ensures stable and reliable operation of the TDCM beamline
in long experimental measurements and enables flawless switching between XUV
photon energies in the course of experiments. The demonstrated alignment
strategy can be applied in implementations of other XUV optical schemes,
which use conical diffraction.

The TDCM beamline in combination with a velocity map imaging spectrometer 
is used to investigate generation of IR-assisted sidebands in the XUV 
photoionization of argon atoms. The atoms are ionized by XUV pulses in 
presence of IR pulses with variable pulse energy. The number of observed 
sidebands and their intensity depend on the strength of the electric field, 
which is used to determine the intensity of the IR pulses at the focus.

The implementation of TDCM presented here demonstrates
specifications, which are very close to theoretical expectations. The 
transmission in the current design is limited by the coatings of the 
optical elements and the choice of blaze angles for the gratings. The 
setup can thus be optmized for specific wavelength ranges by installing 
gratings with the optimal blaze angle without any further modifications. 
The spectral resolution of the setup is mostly limited by geometric 
aberrations in imaging by toroidal mirrors. If required, it can be 
improved by replacing the toroidal mirrors with parabolic mirrors, which 
however will require a very precise alignment \cite{Frassetto2008}. The 
use of parabolic mirrors can also improve the temporal 
resolution of the TDCM and open a way to implementation of 
wavelength selected attosecond XUV pulses.

\begin{acknowledgments}
We would like to thank Roman Peslin, Reinhard Grosser, Arje Katz, Thomas 
M\"uller, Kathrin Lange and Geert Reitsma for their help in design, 
construction and characterization of the monochromator beamline.
\end{acknowledgments}

\bibliography{MONO}

\end{document}